\documentclass[journal]{IEEEtran}
\usepackage[T1]{fontenc}
\usepackage[utf8]{inputenc}
\usepackage{amsmath,amssymb,amsfonts}
\usepackage{algorithm}
\usepackage{algorithmic}
\usepackage{float}
\usepackage{graphicx}
\usepackage{booktabs}
\usepackage{array}
\usepackage{colortbl}
\usepackage{xcolor}
\usepackage{multirow}
\usepackage{textcomp}
\usepackage{url}
\usepackage{cite}
\usepackage[colorlinks=true,
            linkcolor=blue,
            citecolor=teal,
            urlcolor=magenta]{hyperref}
\usepackage{placeins}
\newcommand{\best}[1]{\cellcolor{gray!25}\textbf{#1}}
\makeatletter
\let\sccg@IEEEtitlepagestyle\ps@IEEEtitlepagestyle
\def\ps@IEEEtitlepagestyle{%
  \sccg@IEEEtitlepagestyle
  \def\@oddfoot{\parbox{\textwidth}{\centering\footnotesize
    This work has been submitted to the IEEE for possible publication.\\
    Copyright may be transferred without notice, after which this version may no longer be accessible.}}%
  \let\@evenfoot\@oddfoot
}
\makeatother

\begin{document}

\title{SCCGBench: Executability-Oriented Evaluation for Real-World Web Service-Call Code Generation}

\author{Shuo~Wang, Guodong~Fan, Ming~Zhu, Jing~Li, and Chun~Wang%
\thanks{Shuo Wang, Ming Zhu, and Jing Li are with Shandong University of Technology, Zibo, China 
(e-mail: 25405010689@stumail.sdut.edu.cn; zhu\_ming@sdut.edu.cn; li\_jing@sdut.edu.cn).}%
\thanks{Guodong Fan is with Shandong Agriculture and Engineering University, Zibo, China 
(e-mail: guodong.fan@126.com).}%
\thanks{Chun Wang is with Concordia University, Montreal, QC, Canada 
(e-mail: chun.wang@concordia.ca).}%
\thanks{Corresponding author: Jing Li.}%
}

\maketitle

\begin{abstract}
LLM-generated service-call code is increasingly important for service computing, where developers rely on large language models (LLMs) to compose services and invoke third-party services.
However, current approaches have not yet evaluated the invocation and execution of services in the actual code environment.
To support execution-grounded evaluation, we present the Service-Call Code Generation Benchmark (SCCGBench), a benchmark of 3{,}135 traceable and decontamination-filtered RapidAPI call functions from real GitHub repositories. SCCGBench uses redacted-intent generation to reduce circular leakage and introduces Mock, Robust, ParamF1, and host-, endpoint-, method-, and parameter-level diagnostics. Mock denotes a virtual-execution hit that simultaneously matches the gold host, endpoint, and HTTP method, while Robust further requires status checking and exception handling. We evaluate six code LLMs and one non-neural retrieval baseline under intent-only, retrieval-augmented, and gold-documentation Oracle settings. Results show that models often generate syntactically correct or superficially similar HTTP-request code but achieve low Mock success in the intent-only setting; the strongest evaluated model obtains only $3.76\%$ Mock. Under the gold-documentation Oracle settings (C3 and C5), Mock reaches $88.73\%$--$90.52\%$ for the open-source models evaluated in RQ2 and $76.47\%$--$81.86\%$ for DeepSeek-V4-Pro, whereas sparse and dense retrieval remain far below this oracle level.  The findings indicate that service/API grounding is the main bottleneck: Mock is dominated by host, endpoint, and method alignment, while ParamF1 separately reveals parameter-alignment errors. SCCGBench provides an execution-grounded benchmark for Web service invocation, API recommendation, service-interface retrieval, endpoint grounding, and reliable service-call automation.
\end{abstract}

\begin{IEEEkeywords}
Service computing, Web service invocation, API recommendation, code generation benchmark, large language models, executability evaluation, data contamination.
\end{IEEEkeywords}

\section{Introduction}
\label{sec:intro}

Service computing increasingly relies on software agents and large language models (LLMs) that can transform user intents into executable interactions with external Web services~\cite{2026LLMs}. In a typical service-oriented application, a developer may ask an assistant to retrieve weather information, query financial data, translate text, or access travel services through APIs. The generated code must not only be syntactically valid; it must also select the correct service host, endpoint route, HTTP method, request headers, and query/body parameters. Therefore, service-call code generation is a service-interface grounding problem as well as a code synthesis problem.

\begin{figure*}[!t]
  \centering
  \includegraphics[width=0.92\textwidth]{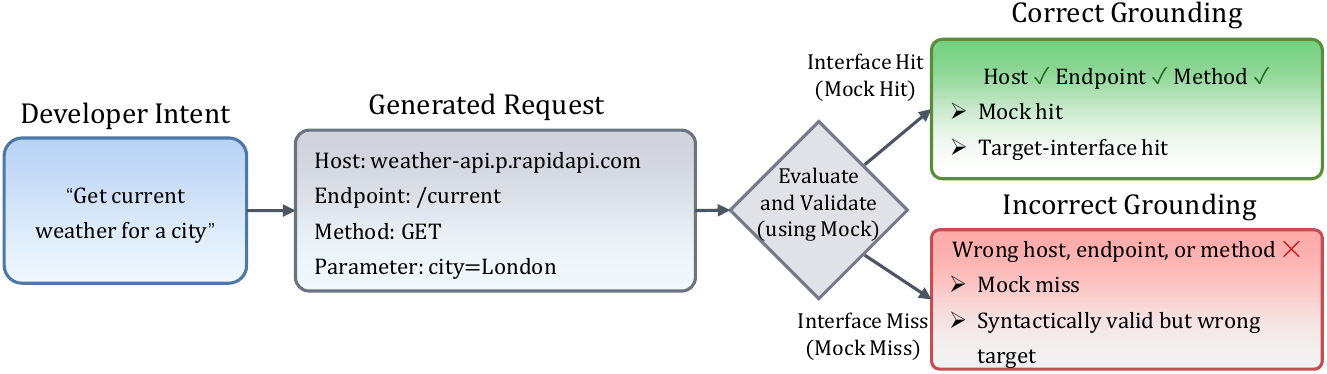}
  \caption{Motivating example of service-interface grounding in service-call code generation.}
  \label{fig:fig1}
\end{figure*}

As a motivating example, consider the developer intent ``get the current weather for a city.'' The model must correctly select the host of the target service, e.g., \texttt{weather-api.p.rapidapi.com}, the endpoint route, e.g., \texttt{/current}, the HTTP method, and valid query parameters, such as the city name. Fig.~\ref{fig:fig1} illustrates this grounding requirement: a service-call program is executable only when the generated request is aligned with the intended service interface rather than merely resembling plausible HTTP-request code. If either the host or the endpoint points to an incorrect target, for example by invoking another weather service or misspelling the route, the generated code fails to hit the intended interface even if it is syntactically complete and structurally plausible. In engineering terms, the entire request fails. As our experiments show (Section~\ref{sec:results}), current models most often fail exactly at this stage of selecting the right service: the vast majority of failures are caused by incorrect hosts rather than by an inability to construct requests. This highlights that evaluating service-call code generation must go beyond textual similarity and directly measure the \emph{executability} of requests.

The community has proposed a series of benchmarks for evaluating code models. HumanEval and MBPP evaluate algorithmic correctness using standalone functions and unit tests~\cite{chen2021codex,austin2021mbpp}; CrossCodeEval and CoderEval introduce cross-file context to better approximate repository-level engineering~\cite{ding2023crosscodeeval,yu2024codereval}; and ComplexCodeEval further emphasizes high-star repositories, real dependencies, and timestamp-based designs to mitigate data leakage~\cite{feng2024complexcodeeval}. These benchmarks have made important progress in algorithmic correctness, contextual dependency, and realism. However, directly transferring them to Web service-call evaluation leaves three limitations insufficiently addressed.

First, tasks and metrics are misaligned with service computing. Existing benchmarks usually rely on static similarity measures such as CodeBLEU and Edit Similarity or on unit-test pass rates. Static similarity can indicate whether generated code looks like a reference implementation, but it cannot determine whether a request hits the correct service endpoint. Unit-test evaluation is also difficult to apply at scale because real Web service calls depend on credentials, network reachability, billing, rate limits, and service availability. Second, sample realism and traceability remain insufficient. Many public data sources contain automatically generated wrappers or tool-learning templates whose signatures and request structures are highly repetitive. Such samples do not reflect the diversity of in-the-wild API usage in real repositories. Third, data contamination and circular leakage remain a concern. When missing comments are completed by asking an LLM to directly transcribe code, the resulting comments may leak hosts, endpoints, parameter keys, or URL fragments, thereby turning interface grounding into answer extraction.

To address these issues, this study presents the Service-Call Code Generation Benchmark (SCCGBench), an executability-oriented benchmark for real-world Web service-call code generation. SCCGBench mines RapidAPI call functions from real GitHub repositories and semantically aligns them with official documentation. It detects and removes template and automatically generated wrapper code, preserves complete provenance information such as repository, file URL, stars, license, and update time, and uses \emph{redacted intent} generation for samples with missing comments. Specifically, before the code is given to the generation model, hosts, URLs, endpoints, and parameter keys/values are masked, and only the \emph{user goal} is described in the style of real developer intents. Paired validation and group-wise ablations show no evidence that redacted intents provide an execution advantage over human comments, supporting the effectiveness of the decontamination procedure.

Based on SCCGBench, we conduct a systematic evaluation of one non-neural baseline and six code LLMs around three research questions. RQ1 investigates whether static similarity is consistent with real executability. RQ2 examines whether retrieval augmentation and gold documentation can overcome the bottleneck and where the bottleneck lies. RQ3 evaluates the reliability of the benchmark with respect to data contamination, temporal leakage, metric trustworthiness, and decoding diversity. The experiments yield three main findings. First, models ``look right but fail to execute'': their Mock scores remain below $3.8\%$ in the intent-only setting, including a frontier reasoning-capable model evaluated in non-think mode. Second, API documentation retrieval and endpoint grounding form the core bottleneck: with gold documentation, Mock scores increase to approximately $89$--$90\%$ for most open-source models and about $82\%$ for DeepSeek-V4-Pro, whereas BM25 and dense retrieval still leave many correct endpoints outside the candidate set. Third, decontamination ablation, temporal slicing, human--machine agreement, decoding-diversity analysis, and bounded real-call validation jointly support the reliability of the data and evaluation.

The main contributions of this paper are as follows:
\begin{itemize}
  \item \textbf{Benchmark contribution}: We propose SCCGBench, a provenance-preserving benchmark for real-world Web service-call code generation. It contains 3{,}135 samples, 850 APIs, and 2{,}082 repositories, and uses redacted intent generation with paired validation to substantially reduce data contamination and circular leakage.
  \item \textbf{Evaluation-framework contribution}: We design an executability-oriented evaluation framework for service-interface grounding. It includes Mock/Robust metrics, host/endpoint/method/parameter-level diagnostics, and a macro--micro analysis strategy that connects generated code to service invocation correctness.
  \item \textbf{Service-computing contribution}: We identify API documentation retrieval and endpoint grounding as the central bottlenecks for reliable Web service invocation, quantifying the gap between intent-only generation, RAG, and gold-documentation oracle settings. This provides direct implications for API recommendation, service discovery, and retrieve--select--generate service-call automation.
  \item \textbf{Reliability contribution}: We validate benchmark reliability through comment-contamination analysis, temporal slicing, human--machine agreement, decoding-diversity analysis, and bounded real-call validation, while explicitly discussing the limitations of virtual execution, Python-only data, and the RapidAPI ecosystem.
\end{itemize}

\section{Background and Related Work}
\label{sec:related}
Service computing studies the modeling, discovery, composition, management, and reliable invocation of network-accessible services. Web service invocation requires binding to a concrete service host, route, method, and parameter schema. Traditional service recommendation and mashup studies have modeled service selection through dynamic service-bundle recommendation and, more recently, multi-agent collaborative reasoning~\cite{liu2023dysr,liu2026mars}. The LLM-based task studied here shifts this binding process from human-written code to model-generated code, making it closely related to Web service discovery, API recommendation, service selection, and service management.

Large code models have improved general code synthesis through pretrained and instruction-aligned representations~\cite{feng2020codebert,guo2021graphcodebert,wang2021codet5,roziere2023codellama,li2023starcoder,guo2024deepseekcoder,team2024codegemma,hui2024qwen25coder}. Yet service-call code generation depends on external interface knowledge: the model must identify the target host and endpoint from a large and changing API space and then construct a request that satisfies the selected interface. Recent tool/API studies, such as ToolLLM, API-Bank, RestGPT, and Gorilla, show the importance of tool selection and documentation grounding~\cite{qin2023toolllm,li2023apibank,song2023restgpt,patil2023gorilla}; SCCGBench further evaluates complete executable request construction at the code level.

Existing code benchmarks cover algorithmic tasks, repository-level completion, and real-world issue resolution~\cite{chen2021codex,austin2021mbpp,hendrycks2021apps,ding2023crosscodeeval,yu2024codereval,jimenez2024swebench}. More recent work improves realism and contamination control through high-quality repositories, evolving tasks, or timestamp-based designs~\cite{feng2024complexcodeeval,jain2024livecodebench}. These benchmarks are valuable, but they rarely evaluate whether an HTTP request hits the correct external service endpoint, and they do not address circular leakage caused by LLM-completed comments that may transcribe hosts or routes.

\begin{table*}[!t]
  \centering
  \caption{Comparison with Existing Benchmarks}
  \label{tab:related}
  \small
  \setlength{\tabcolsep}{4pt}
  \resizebox{0.95\textwidth}{!}{%
  \begin{tabular}{cccccc}
    \toprule
    Benchmark & Service call & Executability & Endpoint alignment & Real provenance & Anti-contamination \\
    \midrule
    HumanEval/MBPP & $\times$ & Tests & $\times$ & $\times$ & $\times$ \\
    CrossCodeEval & $\times$ & $\times$ & $\times$ & $\checkmark$ & Partial \\
    CoderEval & $\times$ & Tests & $\times$ & $\checkmark$ & Partial \\
    ComplexCodeEval & Partial & $\times$ & $\times$ & $\checkmark$ & Timestamp \\
    ToolLLM/API-Bank & $\checkmark$ & Partial & $\times$ & Partial & $\times$ \\
    \textbf{SCCGBench} & $\checkmark$ & \textbf{Mock/Robust} & $\checkmark$ & $\checkmark$ & \textbf{Redaction + ablation} \\
    \bottomrule
  \end{tabular}%
  }

  \vspace{1mm}
  \begin{minipage}{0.95\textwidth}
    \footnotesize \emph{Note:} Endpoint alignment indicates whether the benchmark evaluates whether a generated request hits a concrete API endpoint, including host, route, and method.
  \end{minipage}
\end{table*}

Table~\ref{tab:related} summarizes the gap. Existing benchmarks either focus on general functional correctness or evaluate tool selection without complete request executability. SCCGBench addresses this gap through real provenance, redacted intents, and endpoint-level mock execution.

\section{Construction and Evaluation of SCCGBench}
\label{sec:bench}
\begin{figure*}[!t]
  \centering
  \includegraphics[width=\textwidth]{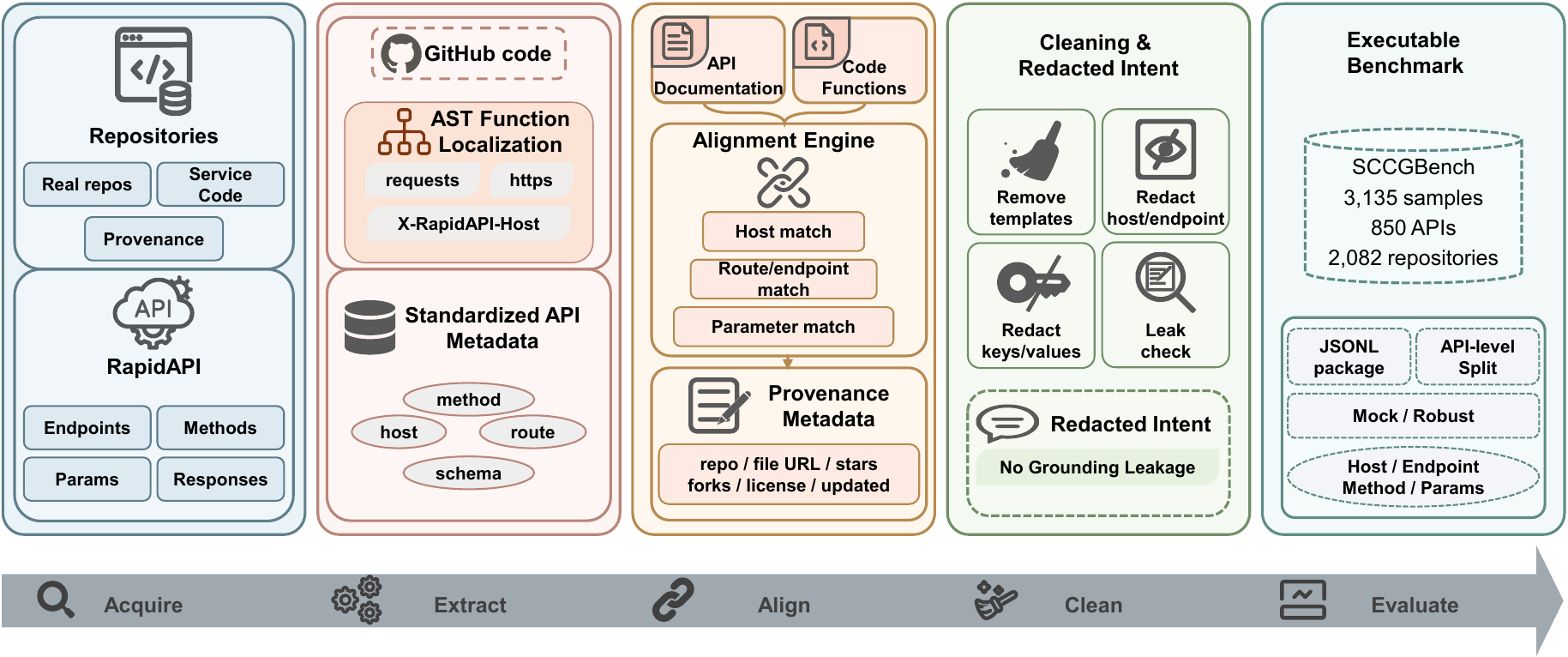}
  \caption{Overview of SCCGBench Construction and Evaluation}
  \label{fig:fig2}
\end{figure*}

This section first formalizes the service-call code generation task. It then presents the overall construction and evaluation pipeline, shown in Fig.~\ref{fig:fig2}. The pipeline covers RapidAPI documentation crawling and GitHub code retrieval, AST-based decoupling, multi-source alignment, template decontamination, redacted intent generation, API-level splitting, and executability evaluation.

\subsection{Problem Formalization}
Given a natural-language intent $q$, and optionally documentation context $c$ in the augmented setting, a model $\mathcal{M}$ is required to generate service-call code $\hat{y}=\mathcal{M}(q,c)$. Each sample is associated with a hidden ground-truth tuple $g=\langle h, e, m, P\rangle$, where $h$ is the API host, $e$ is the endpoint route, $m$ is the HTTP method, and $P$ is the parameter set. Executing $\hat{y}$ in a controlled virtual environment captures the request it attempts to issue, denoted as $r(\hat{y})=\langle \hat h,\hat e,\hat m,\hat P\rangle$. The goal of executability evaluation is to measure the consistency between $r(\hat{y})$ and $g$, rather than the textual similarity between $\hat{y}$ and the reference implementation $y$.

\noindent\emph{Formalization of evaluation metrics.} Let the test set be $\mathcal{D}=\{(q_i,g_i,y_i)\}_{i=1}^{N}$, and let $\mathbb{I}[\cdot]$ denote the indicator function. All interface-oriented metrics are defined as averages over $\mathcal{D}$. The hit accuracy of hosts, endpoints, and methods is defined as
\begin{equation}
\begin{aligned}
\text{Host@Acc} &= \tfrac1N\!\sum_i \mathbb{I}[\hat h_i{=}h_i],\\
\text{Endpoint@Acc} &= \tfrac1N\!\sum_i \mathbb{I}[\hat e_i{=}e_i],
\end{aligned}
\end{equation}
and Method@Acc is defined analogously over $m$. Endpoint@Acc is a route-level, host-independent metric: it measures whether the generated route string matches the gold route, even if the generated host is wrong. Therefore, Endpoint@Acc can be higher than Host@Acc and should not be interpreted as a complete interface hit. ExecReq measures whether $\hat y_i$ can successfully construct a structurally complete HTTP request in the virtual environment. Documentation-parameter F1 (ParamF1) compares the parameters actively generated in the request against the documentation-derived reference parameter set $P_i$, measuring the precision and recall of $\hat P_i$ with respect to $P_i$. The core metric, \emph{mock-execution hit} (Mock), requires the host, endpoint, and method to all match simultaneously and is therefore the final service-interface hit metric:
\begin{equation}
\label{eq:mock}
\text{Mock}=\tfrac1N\!\sum_i \mathbb{I}\!\left[\hat h_i{=}h_i \wedge \hat e_i{=}e_i \wedge \hat m_i{=}m_i\right].
\end{equation}
The robust call rate (Robust) further requires that the generated code contains status checking and exception handling in addition to satisfying Mock, thereby measuring whether the request is both correct and robust:
\begin{equation}
\label{eq:robust}
\text{Robust}=\tfrac1N\!\sum_i \mathbb{I}\!\left[\text{Mock}_i \wedge \textsc{HasErrHandling}(\hat y_i)\right].
\end{equation}
Intuitively, ExecReq measures whether a request can be constructed; Host@Acc, Endpoint@Acc, and Method@Acc provide component-level grounding diagnostics, Mock evaluates host, endpoint, and method matching, and ParamF1 reports parameter alignment separately. Because Host@Acc and Endpoint@Acc are computed independently, the conditional endpoint accuracy after a correct host is further analyzed in the failure-attribution results. This family of metrics enables precise attribution of failures to different stages of interface grounding (Section~\ref{sec:results}).

\subsection{Data Acquisition and Multi-Source Alignment}
\noindent\emph{Documentation-side acquisition.} We use automated tools to crawl endpoint descriptions, parameter schemas, and response definitions from RapidAPI, and normalize the documentation metadata with reference to the OpenAPI specification. The metadata for each API includes its host, the routes of its endpoints, HTTP methods, parameter keys, and type constraints. These metadata serve as the ground truth for subsequent alignment and evaluation.

\noindent\emph{Multi-level code-side retrieval.} On the code side, we retrieve relevant repositories from GitHub~\cite{kalliamvakou2014github} and locate functions that truly contain request logic through AST parsing. To balance recall and precision, we adopt a \emph{multi-level retrieval strategy from strict to loose matching}: exact endpoint URL matching, route+host matching, host matching, characteristic request-header matching such as \texttt{x-rapidapi-host}, request-library matching such as \texttt{requests/httpx}, and function-definition matching. We also use service-specific features such as the \texttt{*.p.rapidapi.com} domain to precisely extract code snippets reflecting real application scenarios. Earlier retrieval levels provide stronger endpoint-grounding evidence, while later levels improve coverage for long-tail interfaces.

\noindent\emph{Multi-source alignment and traceability.} We then semantically align service identifiers with code snippets, mapping interface definitions to real implementations. According to the strength of the alignment evidence, we annotate the \emph{documentation matching granularity} (doc\_match\_type) as \emph{url} for a complete endpoint URL match, \emph{route} for route+host matching, or \emph{host\_only} for host-only matching. This hierarchy records the amount of grounding information available for each sample and provides a natural stratification dimension for micro-level diagnostics (Section~\ref{sec:results}). We also record repository metadata, including \texttt{repository}, \texttt{file\_url}, stars, forks, license, and latest update time, making each sample \emph{traceable}. Cross-repository deduplication is performed using code hashes to avoid counting the same implementation multiple times.

\subsection{Data Processing and Decontamination}
To ensure quality and realism, we perform multi-dimensional cleaning and semantic enhancement.

\noindent\emph{(1) AST-based function decoupling and validation.} We use ASTs to accurately locate and extract target request functions, remove irrelevant contextual redundancy, perform static syntax validation, and format the code according to conventions such as PEP\,8 to improve atomicity and readability.

\noindent\emph{(2) Decontamination of templates and wrapper code.} Early data contained many automatically generated templates and tool-learning wrapper programs, which can be recognized by signatures such as fixed placeholder key parameters and highly similar request structures. The ``comments'' of these samples are also highly repetitive placeholders, such as ``Define the target RapidAPI endpoint URL,'' and lack real intent. We \emph{filter out detected} template-style samples according to these signatures and follow-up audits, so the final dataset is not dominated by automatically generated wrapper data.

\noindent\emph{(3) Redacted intent generation, the core design of this work.} For samples without valid comments, a naive approach is to ask an LLM to read the code and generate a comment. However, this may \emph{transcribe} the request construction line by line and place answer information such as the host, endpoint, and parameters into the input, causing circular leakage. We therefore use redacted intent generation with three steps. \emph{(a) Redaction}: using the ground-truth metadata $g$, we replace the host, full URL, endpoint path, parameter keys/values, and keys in the code with placeholders to obtain de-identified code $\tilde y$. \emph{(b) Style-guided generation}: a small number of real human comments $E$ are used as few-shot style anchors, and the model is asked to describe only the \emph{user goal} based on $\tilde y$, such as ``translate a piece of text into another language,'' rather than restating implementation details. \emph{(c) Leakage checking and retry}: $\textsc{LeakCheck}(t,g)$ determines whether the generated comment $t$ still contains identifying tokens in $g$, such as the host, endpoint path, or parameter keys. If it does, we retry once with a stricter prompt. The resulting comments are natural and close to human style, but contain no answer grounding. This preserves task semantics while substantially reducing the shortcut of reading the answer directly from the comment. Section~\ref{sec:rq3} reports paired validation showing no statistically significant executability advantage for the generated intents over real human comments.

\noindent\emph{(4) Dependency and parameter-schema extraction.} We traverse AST \texttt{import} nodes to automatically identify third-party dependencies required for execution, such as \texttt{requests}. We also extract fine-grained parameter definitions, including keys, locations, and example values, from function signatures, request URLs, \texttt{headers}, \texttt{params}, and request-body construction. The extracted information serves as ground truth for parameter-level correctness validation (ParamF1) and is aligned with the documentation parameter set $P$, allowing the evaluation to refer to official documentation parameters and to fall back to parameters observed in real call code when documentation is missing.

We use \emph{paired experiments} to validate the non-leakage property of redacted intents. On the same set of 291 human-commented samples, we use human comments and redacted intents as inputs and compare the generated code of the strongest model. The paired permutation tests yield $p{=}0.68$ for Mock and $p{=}0.65$ for ES, both non-significant. As a contrast, if unredacted comments directly transcribed from the code are used, the task becomes significantly easier, with Mock $7.9$ vs. $3.4$ and $p{=}0.012$. This quantitative result supports that the redaction mechanism substantially mitigates circular leakage; details are given in Section~\ref{sec:rq3}.

\subsection{Data Packaging, Quality, and Statistics}
All samples are packaged as standard JSON/JSONL. Each follows an ``intent input--implementation output'' structure and preserves multidimensional metadata, including programming language, third-party dependencies, function signature, the aligned API tuple $g=\langle h,e,m,P\rangle$, documentation matching granularity, repository social signals (stars/forks/license/update time), and comment source. Comment sources are divided into three categories: \emph{original} comments from real repositories, \emph{human-style intents}, and \emph{strictly redacted intents}; the latter two are generated by the redacted intent procedure and differ in their degree of style polishing. This packaging allows each sample to be used for evaluation while also supporting provenance review.

\noindent\emph{Quality and realism.} Instead of relying on a single star threshold, we characterize quality using \emph{traceability, real-repository origin, and maintenance activity}. The data come from 2{,}082 distinct repositories, with average stars and forks of 35.83 and 8.39, respectively. After removing template and automatically generated dataset-style repositories, samples are more dispersed across real application projects, and the share of any single repository is very low, preventing a small number of large repositories from dominating the dataset. We also manually audit 50 complex samples and confirm that the precise mapping between RapidAPI endpoint/parameter documentation schemas and AST-extracted code blocks reaches 96\%.

\noindent\emph{Complexity and dependencies.} The samples have an average of 36.2 effective lines of code and a median of 28, reflecting real business complexity. Regarding dependencies, \texttt{requests} appears in 84.8\% of samples. Other frequent dependencies include \texttt{os} (42.5\%), \texttt{django} (38.9\%), \texttt{json} (36.9\%), and \texttt{datetime} (21.4\%), reflecting real engineering contexts.

\noindent\emph{Freshness, split, and coverage.} Repository updates are concentrated in 2024--2026 (about 79\%), and we adopt an API-level split so that the same API does not appear across train/validation/test. Overall statistics are shown in Table~\ref{tab:dataset}. The dataset covers 850 distinct APIs with a long-tail distribution: 418 APIs appear only once and the top 20\% contribute 64\% of samples. RapidAPI metadata shows that about 71\% of samples come from services with at least 1{,}000 subscriptions and 84\% correspond to FREE/FREEMIUM interfaces, supporting reproduction.

\begin{table}[!t]
  \centering
  \caption{Dataset Statistics}
  \label{tab:dataset}
  \small
  \setlength{\tabcolsep}{3pt}
  \renewcommand{\arraystretch}{1.08}
  \begin{tabular}{>{\centering\arraybackslash}m{0.64\linewidth}>{\centering\arraybackslash}m{0.28\linewidth}}
    \toprule
    Statistic & Value \\
    \midrule
    Total samples & 3{,}135 \\
    \multicolumn{1}{c}{Train / validation / test} & 2{,}257 / 266 / 612 \\
    Number of distinct APIs & 850 \\
    Number of distinct repositories & 2{,}082 \\
    Average effective LOC / median & 36.2 / 28 \\
    Real human comments (original) & 1{,}726 \\
    Redacted intent comments (human-style / strict redaction) & 1{,}409 (1{,}263 / 146) \\
    Template / automatically generated wrapper code & 0 \\
    \bottomrule
  \end{tabular}
\end{table}

\section{Experimental Setup}
\label{sec:setup}
Based on SCCGBench, our experiments aim to answer the following research questions:
\begin{itemize}
  \item \emph{RQ1}: When only a natural-language intent is given, how do LLMs perform in terms of static similarity and real executability, and are the two consistent?
  \item \emph{RQ2}: Can retrieval-augmented generation (RAG)~\cite{lewis2020rag,karpukhin2020dpr} and gold-documentation Oracle settings improve executability? Where is the bottleneck?
  \item \emph{RQ3}: Is SCCGBench reliable and fair with respect to data contamination, temporal leakage, automatic metric trustworthiness, and decoding diversity?
\end{itemize}

\subsection{Model Selection}
We evaluate one non-neural retrieval baseline and six code LLMs, listed in Table~\ref{tab:models}. DeepSeek-V4-Pro is evaluated in Non-think mode for protocol consistency; its result should be interpreted as a reasoning-capable model under a non-think inference setting. BM25-nearest-train retrieves the most similar training intent and directly copies its real code. Because the API-level split is not a raw host-string split, related services can share a host; therefore BM25 may occasionally match a host but still fail endpoint/method matching, and its Mock remains zero. Local models are deployed via Ollama~\cite{guo2024deepseekcoder,team2024codegemma,hui2024qwen25coder}; all models use temperature $0$, top-$p{=}1$, and one greedy generation.

\begin{table}[htbp]
  \centering
  \caption{Evaluated Models}
  \label{tab:models}
  \small
  \setlength{\tabcolsep}{3pt}
  \renewcommand{\arraystretch}{1.08}
  \begin{tabular}{>{\centering\arraybackslash}m{0.48\linewidth}>{\centering\arraybackslash}m{0.22\linewidth}>{\centering\arraybackslash}m{0.22\linewidth}}
    \toprule
    Model & Scale & Type \\
    \midrule
    BM25-nearest-train & --- & Retrieval baseline \\
    DeepSeek-Coder-1.3B & 1.3B & Base \\
    Qwen2.5-Coder-7B & 7B & Instruct \\
    CodeGemma-7B-Instruct & 7B & Instruct \\
    Qwen2.5-Coder-32B & 32B & Instruct \\
    DeepSeek-Coder-33B & 33B & Base \\
    DeepSeek-V4-Pro~\cite{deepseek2026v4pro} & Frontier (1.6T/49B activated) & Reasoning-capable \\
    \bottomrule
  \end{tabular}
\end{table}

\subsection{Task Design and Prompt Template}
Given a natural-language intent, and in augmented settings the provided documentation or candidates, a model must complete two steps. The first is parameter filling: supplying valid values according to parameter types and constraints. The second is executable code synthesis: generating a complete Python function with correct dependency imports, URL/header/parameter construction, a \texttt{requests} call, and response parsing. For standardized evaluation, all intent-only, RAG, and Oracle C3 settings share the same core prompt: the model is asked to act as an experienced Python engineer, use placeholder keys, generate only code without explanations or Markdown markers, and construct a complete request according to the available context. The prompt does not impose a fixed error-handling template in these settings. Oracle C5 is an additional robustness-oriented setting: it provides the same gold interface grounding as C3 but explicitly adds robustness constraints such as status checking and \texttt{try-except} exception handling. This distinction explains why C5 can substantially improve Robust after the correct endpoint has already been grounded, while C3 mainly measures whether gold documentation solves the service-interface grounding problem.

\subsection{Evaluation Metrics}
\emph{Static similarity}: CodeBLEU~\cite{ren2020codebleu}, which combines n-gram, AST, and data-flow matching; character-level normalized Levenshtein Edit Similarity (ES); and token-level Precision/Recall/F1 measuring alignment of elements such as endpoints and key parameters. \emph{Executability diagnostics}: Syntax correctness, dependency validity (Dep.), secret safety (Secret), Host@Acc, Endpoint@Acc, Method@Acc, documentation-parameter F1 (ParamF1), request constructability (ExecReq), mock-execution hit rate (Mock), and robust call rate with exception handling (Robust). Mock is defined in Section~\ref{sec:bench}. \emph{Retrieval quality}: Host@K and Endpoint@K, where $K{\in}\{1,3,5,10\}$. \emph{Robustness}: group-wise metric differences, permutation-test $p$-values, and Cohen's $\kappa$~\cite{artstein2008agreement}.

\subsection{Virtual Execution Environment and Mock Judgment}
To evaluate whether a request hits the correct interface without relying on real networks or credentials, we implement a lightweight \emph{virtual execution environment}. Given generated code $\hat y$, we first parse it into an AST and execute it in an isolated namespace. Before execution, we replace the real network libraries with fake \texttt{requests} and \texttt{httpx} modules through \texttt{sys.modules} injection. The fake modules intercept calls such as \texttt{get/post/put/delete/patch/request}, capture the target URL, HTTP method, request headers, and parameters, and return a placeholder response without issuing any real network request. From the captured request, we parse $\langle\hat h,\hat e,\hat m,\hat P\rangle$ and determine Mock and Robust according to Eqs.~(\ref{eq:mock}) and~(\ref{eq:robust}). Execution is protected with timeouts and exception handling so that infinite loops or runtime errors in individual samples do not affect the overall evaluation. After execution, the original modules are restored to avoid side effects on the evaluation process. This design makes executability evaluation close to real call semantics while remaining large-scale, deterministic, reproducible, free of API quota consumption, and free of real billing.

\subsection{Evaluation Strategy: From Macro to Micro}
To approximate real developer interaction, an autonomous agent faces natural-language queries whose complexity is unknown and cannot rely on prior knowledge of the target codebase. We therefore first perform \emph{macro-level} aggregate evaluation over all instances to establish a strict global baseline. We then perform \emph{micro-level} group diagnostics, such as by documentation matching type, comment source, and temporal slice, to reveal capability boundaries under different conditions while isolating sources of variation.

\section{Experimental Results}
\label{sec:results}
We answer the three research questions on the test set of $612$ samples. Overall, the experiments yield three mutually supporting findings. First, static similarity diverges sharply from real executability: models ``look right but fail to execute,'' corresponding to RQ1. Second, providing gold documentation improves executability by an order of magnitude, supporting the conclusion that API documentation retrieval and endpoint grounding are the core bottlenecks, corresponding to RQ2. Third, decontamination, temporal analysis, human--machine agreement, and decoding-diversity analysis jointly support the effectiveness of the benchmark decontamination procedure and the reliability of the metrics, corresponding to RQ3. We discuss these findings in turn.

\subsection{RQ1: Benchmark Performance and Executability Diagnosis}
\emph{Purpose.} We examine whether static similarity can represent real executability.

\emph{Results.} Table~\ref{tab:rq1-sim} reports static similarity. Consistent with typical observations in code generation benchmarks, the BM25 retrieval baseline that copies the nearest real training code achieves the highest CodeBLEU/ES/Recall, while neural models are lower. The reasoning-capable DeepSeek-V4-Pro achieves a CodeBLEU of 25.18, second only to BM25. Notably, neural models show a \emph{high-Precision, low-Recall} pattern. For example, Qwen-32B has Token Precision 66.75 and Recall 40.52, suggesting that its outputs are concise but do not fully cover the reference implementation. The smallest model, DeepSeek-Coder-1.3B, obtains the lowest scores, consistent with its syntax correctness of only 31.9\%.

\begin{table}[htbp]
  \centering
  \caption{RQ1: Static Similarity Results}
  \label{tab:rq1-sim}
  \small
  \setlength{\tabcolsep}{3pt}
  \begin{tabular}{cccccc}
    \toprule
    Model & CodeBLEU & ES & Prec. & Recall & F1 \\
    \midrule
    BM25 & \best{28.23} & \best{29.95} & 52.22 & \best{50.85} & 46.29 \\
    DS-Coder-1.3B & 12.64 & 16.15 & 34.65 & 31.40 & 24.47 \\
    Qwen2.5-7B & 18.91 & 27.16 & 65.93 & 41.00 & 47.38 \\
    CodeGemma-7B & 18.35 & 26.00 & 65.48 & 36.61 & 43.60 \\
    Qwen2.5-32B & 19.00 & 28.04 & \best{66.75} & 40.52 & 47.42 \\
    DS-Coder-33B & 21.56 & 25.89 & 60.87 & 43.18 & 46.66 \\
    DS-V4-Pro & 25.18 & 27.42 & 55.30 & 48.25 & \best{47.47} \\
    \bottomrule
  \end{tabular}

  \vspace{1mm}
  \begin{minipage}{\linewidth}
    \footnotesize \emph{Note:} Results are reported on the test set ($n=612$). The best value in each column is shaded and boldfaced.
  \end{minipage}
\end{table}

\begin{table*}[!t]
  \centering
  \caption{RQ1: Executability Diagnostics}
  \label{tab:rq1-diag}
  \small
  \setlength{\tabcolsep}{3pt}
  \renewcommand{\arraystretch}{1.1}
  \resizebox{0.95\textwidth}{!}{%
  \begin{tabular}{ccccccccccc}
    \toprule
    Model & Syntax & Dep. & Secret & Host@Acc & Endpoint@Acc & Method@Acc & ParamF1 & ExecReq & Mock & Robust \\
    \midrule
    BM25-nearest-train & 99.67 & 94.44 & 56.37 & 5.72 & 8.82 & 82.84 & 16.19 & 15.20 & 0.00 & 0.00 \\
    DeepSeek-Coder-1.3B & 31.86 & 72.55 & 99.67 & 2.29 & 9.80 & 60.62 & 4.82 & 9.80 & 0.00 & 0.00 \\
    Qwen2.5-Coder-7B & \best{100} & 99.35 & 99.84 & 3.10 & 15.03 & 85.78 & 20.80 & 87.25 & 0.65 & 0.49 \\
    CodeGemma-7B & \best{100} & \best{99.67}& 99.35 & 5.56 & 12.09 & 87.09 & 9.95 & \best{91.67} & 1.31 & 1.14 \\
    Qwen2.5-Coder-32B & \best{100} & 99.02 & \best{100} & 6.86 & \best{19.12} & \best{89.38} & 21.71 & 88.56 & 2.94 & 1.31 \\
    DeepSeek-Coder-33B & 99.35 & \best{99.67}& \best{100} & 7.52 & 14.54 & 88.56 & 19.73 & 89.22 & 2.94 & 1.14 \\
    DeepSeek-V4-Pro & 97.55 & 97.88 & 99.84 & \best{8.17} & 18.46 & 87.58 & \best{25.09} & 80.56 & \best{3.76} & \best{3.76} \\
    \bottomrule
  \end{tabular}
  }

  \vspace{1mm}
  \begin{minipage}{0.95\textwidth}
    \footnotesize \emph{Note:} Values are percentages on the test set ($n=612$). The best value in each column is shaded and boldfaced.
  \end{minipage}
\end{table*}

\emph{Observations and interpretation.} Static similarity severely overestimates real usability. The executability diagnostics in Table~\ref{tab:rq1-diag} reveal a structural hallucination paradox: except for the 1.3B base model, most larger neural models approach near-perfect syntactic correctness, request constructability of $80$--$92\%$, and near-perfect secret safety, but their Mock scores, which measure simultaneous host, endpoint, and method matching, are almost zero and never exceed $3.8\%$; host accuracies are only $3$--$8\%$, and endpoint accuracies are only $12$--$19\%$ for these larger neural models. A sharp contrast emerges: models often write requests with the right \emph{form} and high ExecReq, but they usually point to the \emph{wrong} service endpoint, resulting in low Endpoint@Acc and low Mock. Even the strongest evaluated model, the reasoning-capable DeepSeek-V4-Pro in non-think mode, achieves only $3.76\%$ Mock. This result argues against attributing the failure only to weak code synthesis ability and instead indicates that the dominant failure arises from inability to ground the intent to the correct API. Moreover, although BM25 obtains the highest static similarity, its Mock remains $0$ because it copies real code from \emph{another} sample whose host often mismatches the current target or whose endpoint/method does not match even when the host is related. This further confirms the gap between ``looking similar'' and ``being executable.''

\noindent\textbf{Finding 1.} High static similarity does not imply service-call executability; models generally look plausible but fail to execute, and the bottleneck lies in API grounding.

\emph{Micro-level diagnosis.} To further isolate the source of failure, we stratify the test set by the \emph{documentation matching granularity} doc\_match\_type: \emph{host\_only}, where only the host is available; \emph{route}, where the route is available; and \emph{url}, where the complete URL is available. We find that as the available grounding information increases from host\_only to route to url, Endpoint@Acc and Mock increase monotonically. For Qwen-32B, for example, Mock rises from $0\%$ in the host\_only stratum to $4.6\%$ in the url stratum, and Endpoint@Acc rises from $0.9\%$ to $24.9\%$, while Method@Acc and ExecReq remain high at approximately $88\%$ across strata. This micro-level comparison again confirms that the executability bottleneck is not whether a request can be constructed, but whether the intent can be grounded to a sufficiently specific and correct endpoint.

\begin{table}[!t]
  \centering
  \caption{RQ1: Hierarchical Failure Attribution}
  \label{tab:errorattr}
  \small
  \setlength{\tabcolsep}{4pt}
  \renewcommand{\arraystretch}{1.03}
  \begin{tabular}{lcc}
    \toprule
    Earliest stage & Count & Share \\
    \midrule
    Syntax error & 436 & 11.9\% \\
    Request construction failure & 500 & 13.6\% \\
    Host error & 2{,}562 & 69.8\% \\
    Endpoint error & 97 & 2.6\% \\
    Method error & 4 & 0.1\% \\
    Mock hit & 70 & 1.9\% \\
    Unclassified residual & 3 & 0.1\% \\
    \bottomrule
  \end{tabular}

  \vspace{0.5mm}
  \begin{minipage}{\linewidth}
    \footnotesize \emph{Note:} The 3{,}672 generations correspond to 612 test samples evaluated across six neural models, excluding the BM25 retrieval baseline; each generation is assigned to the earliest failed stage.
  \end{minipage}
\end{table}

\emph{Failure attribution.} To locate where failures occur, we analyze all $3{,}672$ generations from the six neural models in the intent-only setting and attribute each failure to the earliest failed stage (Table~\ref{tab:errorattr}). We sequentially check syntax, request construction, host, endpoint, and method, and assign each generation to the first failed stage. The results show that failures \emph{mostly occur at the earliest semantic stage of selecting the right service}: host errors alone account for $69.8\%$ and are by far the dominant source of failure. In contrast, syntax errors ($11.9\%$) and request-construction failures ($13.6\%$) together account for only about one quarter, while endpoint or method errors after a correct host are very rare ($2.6\%+0.1\%$). Only a small number of generations reach a complete Mock hit; the unclassified residual cases are retained for bookkeeping and are not interpreted as a substantive failure category. In other words, model failures are rarely due to an inability to write requests; rather, the request points to the wrong service from the beginning. This is consistent with the diagnostics showing much lower host/endpoint accuracy than syntax and constructability, and provides fine-grained quantitative evidence that the bottleneck is API grounding. Even for the strongest model, DeepSeek-V4-Pro, host errors still account for about $73\%$ of its failures, while syntax errors account for only $2.5\%$, further arguing against the ``weak code synthesis'' explanation.

\emph{Representative failure cases.} To illustrate the nature of incorrect-host failures, we examine two representative failures of the strongest model, DeepSeek-V4-Pro, in the intent-only setting. In both cases, syntax and request construction succeed, but the host/endpoint points to the wrong target. \emph{Case 1: falling back to an official public API}. The intent is ``Get current weather for location.'' The correct target is the RapidAPI wrapper \url{open-weather13.p.rapidapi.com/city}, but the model generates a call to \url{api.openweathermap.org/data/2.5/weather}, falling back to the OpenWeatherMap public API that is familiar from pretraining rather than the target RapidAPI wrapper. A similar phenomenon occurs when an intent for retrieving video information is directed to \url{googleapis.com/youtube/v3}. \emph{Case 2: confusing similar RapidAPI services}. The intent is ``Get Google trends using RapidAPI,'' which explicitly mentions RapidAPI. The correct target is \url{google-realtime-trends-data-api.p.rapidapi.com/trends}, but the model calls \url{realtime-google-trends.p.rapidapi.com}, a similarly named but non-target service. These cases corroborate the statistics in Table~\ref{tab:errorattr}: the dominant problem is not that models cannot construct requests, but that they direct requests to the wrong service, either by falling back to common public APIs or by confusing similar interfaces. This is a micro-level manifestation of the endpoint-grounding bottleneck and shows that parametric knowledge alone is insufficient for reliably locating a specific target service.

\subsection{RQ2: Information Augmentation and the Retrieval Bottleneck}
\emph{Purpose.} We investigate whether augmented information can overcome the executability bottleneck and where the bottleneck lies.

\emph{Results.} Table~\ref{tab:rq2} compares four models across a staircase of six settings: intent-only, RAG using BM25, RAG using MiniLM, RAG using bge, and two Oracle settings (C3 and C5). Dense retrieval (MiniLM/bge) provides richer grounding than BM25, and the results show an overall staircase trend from intent-only/RAG settings to Oracle settings, although dense retrievers are not strictly monotonic for every model and metric. For Qwen2.5-Coder-32B, Mock increases from $2.9\%$ in the intent-only setting to $20.3\%$ under BM25 RAG, further to $25.7$--$28.3\%$ under dense RAG, and finally to approximately $89\%$ under Oracle. The reasoning-capable DeepSeek-V4-Pro model follows the same pattern: $3.8\%$ (intent) $\to$ $20.1\%$ (BM25) $\to$ $22.7$--$26.8\%$ (dense) $\to$ $76.5$--$81.9\%$ (Oracle).

\begin{table*}[!t]
  \centering
  \caption{RQ2: Information-Augmentation Staircase}
  \label{tab:rq2}
  \small
  \setlength{\tabcolsep}{3pt}
  \renewcommand{\arraystretch}{1.1}
  \begin{tabular}{ccccccc}
    \toprule
    Model & Setting & Host@Acc & Endpoint@Acc & ParamF1 & Mock & Robust \\
    \midrule
    \multirow{6}{*}{Qwen2.5-Coder-7B}
      & Intent only & 3.10 & 15.03 & 20.80 & 0.65 & 0.49 \\
      & BM25 RAG & 28.59 & 26.96 & 3.82 & 19.28 & 4.74 \\
      & MiniLM RAG & 34.64 & 33.82 & 5.68 & 25.49 & 6.54 \\
      & bge RAG & 35.62 & 34.80 & 4.79 & 25.49 & 6.05 \\
      & Oracle (C3) & 100 & 95.92 & 94.12 & 88.73 & 0.00 \\
      & Oracle (C5) & 100 & 95.75 & 82.54 & 89.38 & 89.38 \\
    \midrule
    \multirow{6}{*}{Qwen2.5-Coder-32B}
      & Intent only & 6.86 & 19.12 & 21.71 & 2.94 & 1.31 \\
      & BM25 RAG & 30.88 & 28.10 & 3.45 & 20.26 & 15.69 \\
      & MiniLM RAG & 38.40 & 36.11 & 4.99 & 28.27 & 22.88 \\
      & bge RAG & 37.75 & 36.27 & 3.82 & 25.65 & 21.08 \\
      & Oracle (C3) & 100 & 96.24 & 97.01 & 89.38 & 0.33 \\
      & Oracle (C5) & 100 & 95.75 & 91.27 & 89.22 & 89.22 \\
    \midrule
    \multirow{6}{*}{DeepSeek-Coder-33B}
      & Intent only & 7.52 & 14.54 & 19.73 & 2.94 & 1.14 \\
      & BM25 RAG & 27.94 & 25.33 & 2.24 & 18.79 & 5.72 \\
      & MiniLM RAG & 34.48 & 32.68 & 3.05 & 25.82 & 7.03 \\
      & bge RAG & 35.46 & 33.17 & 2.64 & 23.86 & 7.52 \\
      & Oracle (C3) & 100 & 95.10 & 93.14 & 90.52 & 0.49 \\
      & Oracle (C5) & 100 & 96.73 & 81.56 & 90.03 & 90.03 \\
    \midrule
    \multirow{6}{*}{DeepSeek-V4-Pro}
      & Intent only & 8.17 & 18.46 & 25.09 & 3.76 & 3.76 \\
      & BM25 RAG & 31.54 & 28.76 & 3.56 & 20.10 & 20.10 \\
      & MiniLM RAG & 39.22 & 37.42 & 3.53 & 22.71 & 22.22 \\
      & bge RAG & 41.18 & 38.40 & 3.95 & 26.80 & 26.31 \\
      & Oracle (C3) & 98.04 & 92.48 & 79.12 & 76.47 & 75.16 \\
      & Oracle (C5) & 98.69 & 93.63 & 74.98 & 81.86 & 81.86 \\
    \bottomrule
  \end{tabular}

  \vspace{1mm}
  \begin{minipage}{0.95\textwidth}
    \footnotesize \emph{Note:} Values are percentages on the test set ($n=612$). BM25 RAG uses sparse retrieval; MiniLM and bge RAG use dense retrieval with all-MiniLM-L6-v2 and bge-base-en-v1.5, respectively. C3 provides gold interface documentation without an explicit robustness template, whereas C5 adds robustness constraints such as status checking and exception handling.
  \end{minipage}
\end{table*}

\begin{table}[!t]
  \centering
  \caption{RQ2: Retrieval Hit Rates}
  \label{tab:retrieval}
  \small
  \setlength{\tabcolsep}{4pt}
  \begin{tabular}{cccccc}
    \toprule
    Retriever & Metric & @1 & @3 & @5 & @10 \\
    \midrule
    \multirow{2}{*}{BM25 (sparse)} & Host@K & 18.14 & 26.80 & 30.72 & 38.56 \\
                                & Endpoint@K & 13.73 & 23.20 & 26.47 & 33.99 \\
    \midrule
    \multirow{2}{*}{Dense MiniLM} & Host@K & \best{27.78}& 40.20 & 47.55 & 57.03 \\
                                  & Endpoint@K & \best{21.41}& 33.50 & 42.97 & 52.45 \\
    \midrule
    \multirow{2}{*}{Dense bge-base} & Host@K & 27.29 & \best{44.93}& \best{52.29}& \best{58.50} \\
                                    & Endpoint@K & 21.24 & \best{36.93}& \best{46.08}& \best{54.41} \\
    \bottomrule
  \end{tabular}

  \vspace{1mm}
  \begin{minipage}{\linewidth}
    \footnotesize \emph{Note:} Values are Host@K / Endpoint@K percentages. All retrievers use the same 850-API documentation library and the same hit criterion. The best value for each metric at each K is boldfaced.
  \end{minipage}
\end{table}

\emph{Observations and interpretation.} Retrieval quality itself is limited. BM25 reaches only $33.99\%$ Endpoint@10 and $38.56\%$ Host@10, explaining why RAG raises Mock only to about $20\%$: under the top-10 retrieval diagnostic, the correct endpoint is absent for roughly two thirds of samples. Dense retrieval improves Endpoint@10 to $52.45$--$54.41\%$ and average Mock to about $25.5\%$, but the gap to Oracle C3 (Mock $86.3\%$ on average) remains about 61 points. Once gold documentation is provided, most open-source models reach about $89$--$90\%$ Mock, showing that models can generate executable requests when the correct interface is supplied. ParamF1 under RAG stays in the low single digits because a wrong endpoint also leads to wrong parameters. C5 improves Robust by adding explicit status-checking and exception-handling constraints after the endpoint has been grounded. Thus dense retrieval mitigates but does not remove the API documentation retrieval and endpoint-grounding bottleneck.

\noindent\textbf{Finding 2.} API documentation retrieval and endpoint grounding are the core bottlenecks in current service-call code generation. Dense retrieval mitigates but does not eliminate this bottleneck. Models already have some ability to generate requests, and the key gap lies in retrieving the correct interface.

\subsection{RQ3: Reliability and Bias Analysis}
\label{sec:rq3}
\emph{Purpose.} We evaluate benchmark reliability from four perspectives: comment contamination, temporal leakage, metric trustworthiness, and decoding diversity.

\emph{(1) Comment-contamination analysis.} We conduct two complementary analyses. First, on the same set of 291 human-commented samples, we compare model outputs when the input is the original human comment versus when the input is the corresponding redacted intent. This matched design uses paired permutation tests. The tests yield $p{=}0.68$ for Mock and $p{=}0.65$ for ES, both non-significant, showing no evidence that redacted intents provide an execution advantage over real human comments. As a contrast, when unredacted code-transcription comments are used, the task becomes significantly easier, with Mock $7.9$ vs. $3.4$ and $p{=}0.012$. This contrast supports the necessity of redaction.

Second, we perform a group-wise analysis on the full test set by comment source, comparing real human comments $A$ with redacted intents $B$. Because these two groups contain different samples, we treat this as a group-wise comparison rather than a paired test. The group-wise permutation tests show no significant Mock difference for all models; the observed values are nearly identical, e.g., Qwen-32B obtains $2.92$ vs. $2.96$, and DeepSeek-V4-Pro obtains $3.8$ vs. $3.7$. These results do not prove the complete absence of contamination, but they provide no evidence that redacted-intent comments create an unfair executability advantage. The group differences in static metrics such as CodeBLEU/ES are more plausibly explained by differences in sample populations, because group $B$ originally lacks human comments and has different code styles, rather than by comment leakage. The key execution metric Mock shows no meaningful difference.

\emph{(2) Temporal leakage.} We split GitHub samples around June 2024 and compare ES and Mock before and after the split. We use this cutoff as a coarse temporal boundary rather than as a guaranteed knowledge cutoff for every evaluated model, especially for newer models such as DeepSeek-V4-Pro. Mock has $p{\approx}1.0$ for all models, and post-cutoff samples are not worse. For DeepSeek-V4-Pro, the post-cutoff Mock is slightly higher, increasing from $3.11$ to $3.99$. Most ES comparisons have $p{>}0.05$. This suggests that we find no evidence of a memorization-driven execution advantage; the temporal split does not show a clear post-cutoff degradation pattern, and the API-constraint-based evaluation framework remains effective on newer samples.

\emph{(3) Human--machine agreement.} Two annotators blindly score 100 pieces of model-generated code and we compare their judgments with automatic metrics. The results are shown in Table~\ref{tab:rq3}. Grounding metrics such as Host/Endpoint/Method achieve human--machine Cohen's $\kappa$ of $0.86$--$1.0$, indicating near-perfect agreement, and inter-annotator $\kappa$ also reaches $0.88$--$1.0$. For Mock, because its definition requires host, endpoint, and method to all be correct, we combine the human judgments of these three elements as the human counterpart of Mock, yielding human--machine $\kappa{=}0.90$. It is worth noting that if annotators are directly asked the broad question of whether the generated code ``can run successfully,'' humans tend to \emph{overestimate} executability from surface cues alone: the human--machine $\kappa$ is only $0.05$ and the raw agreement is $0.55$, even though the two annotators are highly consistent with each other. This itself shows that humans have difficulty judging real executability by visual inspection alone, and therefore supports the necessity of execution-grounded automatic metrics. For the robustness-related error-handling judgment, human--machine agreement is moderate with $\kappa=0.32$, so Robust is treated as a secondary metric.

\begin{table}[htbp]
  \centering
  \caption{RQ3: Human--Machine Agreement}
  \label{tab:rq3}
  \begin{tabular}{ccc}
    \toprule
    Metric & Inter-annotator $\kappa$ & Human--machine $\kappa$ \\
    \midrule
    Host@Acc & 0.94 & 0.96 \\
    Endpoint@Acc & 0.90 & 0.86 \\
    Method@Acc & 1.00 & 1.00 \\
    Mock (= H$\wedge$E$\wedge$M) & --- & 0.90 \\
    Robust & 0.88 & 0.32 \\
    \bottomrule
  \end{tabular}

  \vspace{1mm}
  \begin{minipage}{\linewidth}
    \footnotesize \emph{Note:} Agreement is measured by Cohen's $\kappa$ on 100 annotated generations.
  \end{minipage}
\end{table}

\emph{(4) Decoding diversity (pass@$k$).} The main experiments use single greedy decoding with temperature $0$, where Mock approximates pass@1. To test whether a single decoding path underestimates model capability, we sample 5 outputs for each of the $612$ test samples using Qwen2.5-Coder-7B with temperature $0.8$ and compute pass@$k$ using the unbiased estimator. The results are pass@1 $=0.56\%$, pass@3 $=1.06\%$, and pass@5 $=1.31\%$. Although executability increases monotonically with the number of samples, the absolute value remains below $1.5\%$, in the same order of magnitude as the $0.65\%$ under greedy decoding. This indicates that increasing sampling diversity alone cannot overcome the executability bottleneck. The root cause is failure to ground the intent to the correct interface, rather than the decoding strategy. Additional samples provide only marginal gains, consistent with the limitations of sampling-based self-consistency reasoning~\cite{wang2023selfconsistency} on this task.

\noindent\textbf{Finding 3.} Comment-contamination analysis, temporal slicing, human--machine agreement, and decoding-diversity analysis jointly support the reliability of SCCGBench and the execution-grounded metrics, while also showing that no single analysis can prove complete absence of contamination.

\section{Discussion and Threats to Validity}
\label{sec:discussion}
\emph{Implications.} SCCGBench shows a structural hallucination pattern: larger models often produce correct-looking request code but fail at service-interface grounding. Mock evaluates host, endpoint, and method matching, while ParamF1 separately reports parameter alignment. The large RAG--Oracle gap indicates that reliable service-call automation should be organized as a retrieve--select--generate pipeline, with specialized API recommendation/reranking as an independently optimizable component. This direction is consistent with retrieval- and action-oriented paradigms such as RAG, ReAct, and self-debugging~\cite{lewis2020rag,yao2023react,chen2023selfdebug}.

\emph{Controlled real-call validation.} To check whether virtual Mock corresponds to real reachability, we conduct a bounded safety-filtered sanity check on FREE RapidAPI services using side-effect-free GET requests. Among 75 screened services, 21 gold requests return 2xx; 19 have Mock-positive generations eligible for safe replay. Mock-positive requests return 2xx in $91.3\%$ of effective attempts (95/104; Wilson 95\% CI $84.4$--$95.4\%$), and 16/19 APIs have at least one successful Mock-positive real call. Five of nine eligible Mock-negative requests also return 2xx because they target different but accepting endpoints on the same host, showing that Mock is stricter than a real 2xx response. This supports the positive predictive validity of Mock under reachable-service conditions but not a general equivalence to real execution.

\emph{Threats to validity.} Construct validity is limited because Mock/Robust use virtual execution rather than full real-network success; we mitigate this through human agreement and the bounded real-call check. Internal validity may be affected by comment-source and temporal differences; paired/group-wise contamination analyses and temporal slicing find no execution-level advantage attributable to these factors. External validity is limited to Python and the RapidAPI ecosystem; extending SCCGBench to JavaScript, other API ecosystems, and larger scales remains future work. Conclusion validity could be affected by weak retrieval or greedy decoding; dense retrievers and pass@$k$ analysis show improvements but do not change the main conclusion that endpoint grounding is the bottleneck.

\section{Conclusion}
\label{sec:conclusion}
This paper addresses the limitations of existing benchmarks in evaluating real Web service-call executability and their susceptibility to circular comment contamination by proposing SCCGBench, an executability-oriented benchmark for service-call code generation. SCCGBench consists of 3{,}135 decontamination-filtered and traceable RapidAPI call functions from real GitHub repositories. It uses redacted intent generation and paired validation to reduce circular leakage, and introduces execution-grounded metrics such as Mock and Robust together with a macro--micro diagnostic framework. A systematic evaluation of six code LLMs and one non-neural retrieval baseline shows that models generally ``look plausible but fail to execute'': even a reasoning-capable frontier model evaluated in non-think mode achieves no more than $3.8\%$ Mock in the intent-only setting. API documentation retrieval and endpoint grounding are the core bottlenecks. Mock increases from below $4\%$ in the intent-only setting to about $20\%$ with BM25 RAG, and most open-source models reach about $89$--$90\%$ under Oracle settings, while retrieval hit rates remain limited, with BM25 Endpoint@10 at only $34\%$ and dense retrieval at about $54\%$. Reliability analyses provide no evidence that redacted intents or older samples create an unfair execution advantage, and human--machine agreement reaches $\kappa=0.86$--$0.96$ for grounding metrics. SCCGBench provides an execution-grounded perspective for evaluating service invocation, API recommendation, and reliable service-call automation in service computing. Future work will investigate service-interface-specific retrieval/reranking models, joint optimization of retrieve--select--generate pipelines, and extensions to multiple programming languages and API ecosystems.
\FloatBarrier

\section*{Data and Code Availability}
The SCCGBench dataset, construction pipeline, and evaluation code have been released at \url{https://github.com/SCCGBench/SCCGBench}, including train/validation/test splits, data-cleaning and redacted-intent scripts, virtual execution, dense retrieval, and pass@$k$ experiments. Third-party credentials are redacted.

\IEEEtriggeratref{18}
\bibliographystyle{IEEEtran}
\bibliography{references}

\end{document}